\documentclass[preprint,12pt,3p]{elsarticle}
\pdfoutput=1

\usepackage{amssymb}

\RequirePackage[OT1]{fontenc}
\RequirePackage{amsthm,amsmath}
\RequirePackage[numbers]{natbib}
\RequirePackage[colorlinks,citecolor=blue,urlcolor=blue]{hyperref}
\usepackage{verbatim}
\usepackage{graphicx}
\usepackage{listings}
\usepackage{hyperref}
\usepackage{subfiles}

\journal{arXiv}

\begin{document}

\begin{frontmatter}
 
\title{Acceptance or Rejection of Lots while Minimizing \\ and Controlling Type I and Type II Errors}
\author[label1]{Edson Luiz Ursini \thanks{corresponding author}}
\ead[url]{https://www.ft.unicamp.br/}

\address[label1]{University of Campinas - School of Technology \\Rua Paschoal Marmo, 1888, Limeira, SP, Brazil}
\author[label1]{Elaine Cristina Catapani Poletti}

\author[label2]{Loreno Menezes da Silveira  }
\author[label1]{José Roberto Emiliano Leite  }

\address[label2]{Teleco Intelligence in Telecommunications\\Avenida Cassiano Ricardo, 601 Salas 63, Sao Jose dos Campos, SP, Brazil} 

\begin{abstract}
The double hypothesis test (DHT) is a test that allows controlling Type I (producer) and Type II (consumer) errors. It is possible to say whether the batch has a defect rate, p, between 1.5 and 2\%, or between 2 and 5\%, or between 5 and 10\%, and so on, until finding a required value for this probability. Using the two probabilities side by side, the Type I error for the lower probability distribution and the Type II error for the higher probability distribution, both can be controlled and minimized. It can be applied in the development or manufacturing process of a batch of components, or in the case of purchasing from a supplier, when the percentage of defects (p) is unknown, considering the technology and/or process available to obtain them. The power of the test is amplified by the joint application of the Limit of Successive Failures (LSF) related to the Renewal Theory. To enable the choice of the most appropriate algorithm for each application. Four distributions are proposed for the Bernoulli event sequence, including their computational efforts: Binomial, Binomial approximated by Poisson, and Binomial approximated by Gaussian (with two variants). Fuzzy logic rules are also applied to facilitate decision-making.
\end{abstract}


\end{frontmatter}


\section{Introduction}
\label{sec1}

In many real-world situations there is no knowledge about the characteristics (for example, failure probability distributions, or number of failures, or interval between failures, etc.) of a product, a component or the process involved to establish the percentage of failures, $p$, associated with them (be it manufactured or purchased). However, we can always associate with them a Bernoulli event, that is, "success" or "failure". The double hypotheses test (DHT) lends itself to the evaluation of the percentage of failures, $p$, associated to this type of events. A typical application is to evaluate the quality of a lot of components already manufactured with or without a specified requirement.
In a less concrete case, in the process of developing or manufacturing a batch of components there is still no idea how many products will be defective in view of the technology and / or process available to obtain them (one can even classify the manufacturing components as being part of a virtual lot, not yet existent). In a more general situation someone can evaluate the quality of service of a telecommunications system. For example, it is stipulated that at most we have a missed call every 100 attempts ($p$ = 1\% failure, which is the call blocking). In practical situations this can occur in real time to avoid congestion. All these situations have in common the fact that events occur according to a distribution of Bernoulli, which may consist of a "success" or a "failure". The proposed DHT can be applied to existing products or lots, or to components that do not yet exist, but are in production, or to evaluate the most diverse types of systems with different thresholds for hypotheses tests. The test is sequential and has a well-defined stopping criterion. Further studies for sequential tests were proposed, as in Anscombe (1953), Hashemi and Pasupathy (2014), and Hashemi et al (2012), but they all had a very large generalization, focusing on very generic distributions and did not address more specifically the issue of Bernoulli's events.  According Louren{\c c}o Filho (1978), there is a sequential acceptance test with Bernoulli events (success or failure), with user-defined acceptance (or rejection) probabilities. Jamkhaneh et al (2009) uses fuzzy logic with some linguistic knowledge to improve estimation of parameter $p$ to facilitate the acceptance or rejection of a lot.  Although simple, our procedure is very important (for large "lots" where the size is not known a priori) because it gives the user an idea of the quality level of the batch being delivered, the manufacturing process under observation, or the QoS (Quality of Service) level of a telephone link. In addition, it also allows control of Type I and Type II errors. Thus, the difference for the proposed procedure is that the DHT works with the two types of error: Type I (producer error or $\alpha$) and Type II (consumer error or $\beta$), but that someone can have control over them since the DHT will always compare one distribution with respect to the other, one of them will be the null hypothesis ($H0$) and the other will be the alternative hypothesis ($Ha$). In practice, if a comparison is made between a distribution with $p_0$ probability and the  other with $p_1$ probability, whatever may be the $H0$ or the $Ha$ hypotheses. Therefore, it is very common to equate the values for errors Type I and Type II (although this is not necessary, depending on the situation). This results the values that will allow to evaluate the acceptance and / or rejection of the lot (or other system under inspection).  Thus, with the two probabilities side by side, the Type I error for the lower probability distribution and the Type II error for the higher distribution can be controlled and minimized. This comparison may continue until the probability of lot defect is found. 
In addition, DHT has its decision-making capacity increased by the fact that our proposal is that it works in conjunction with the Successive Failures Limit (SFL).  The SFL, as in Feller (1968), uses the concepts of Renewal Theory to reject null hypotheses from a number of successive failures (as will be seen in Section 4). However, the major advantage of the procedure is that, although a requirement is not met, the continuity of its implementation may inform that there is another requirement of  lower quality that can be accepted. For instance, a system does not allows 1\% blocking, but the 2\% requirement can be accepted. This is very important when evaluating the system quality. In this way, the test is very useful for evaluating the percentage of defects in a large number of situations where events occur as "success" or "failure". A real-time application might be to evaluate a telecommunications traffic route blocking to allocate more resources to it.  The remainder of this article is organized as follows: Section 2 presents the definitions and concepts related to the DHT. Section 3 shows the application examples and discusses the results obtained. Section 4 deals with the Renewal Theory's concepts on the application of Successive Failures Limit (SFL). Section 5 addresses the main virtues and defects of the methods employed in terms of applications made Finally, Section 6 argues the observations and conclusions.

 \section{Double Hypotheses Test (DHT) for Bernoulli Events}
 \label{sec2}
\begin{figure*}[ht]
     \centering
     \includegraphics[width=0.9\columnwidth]{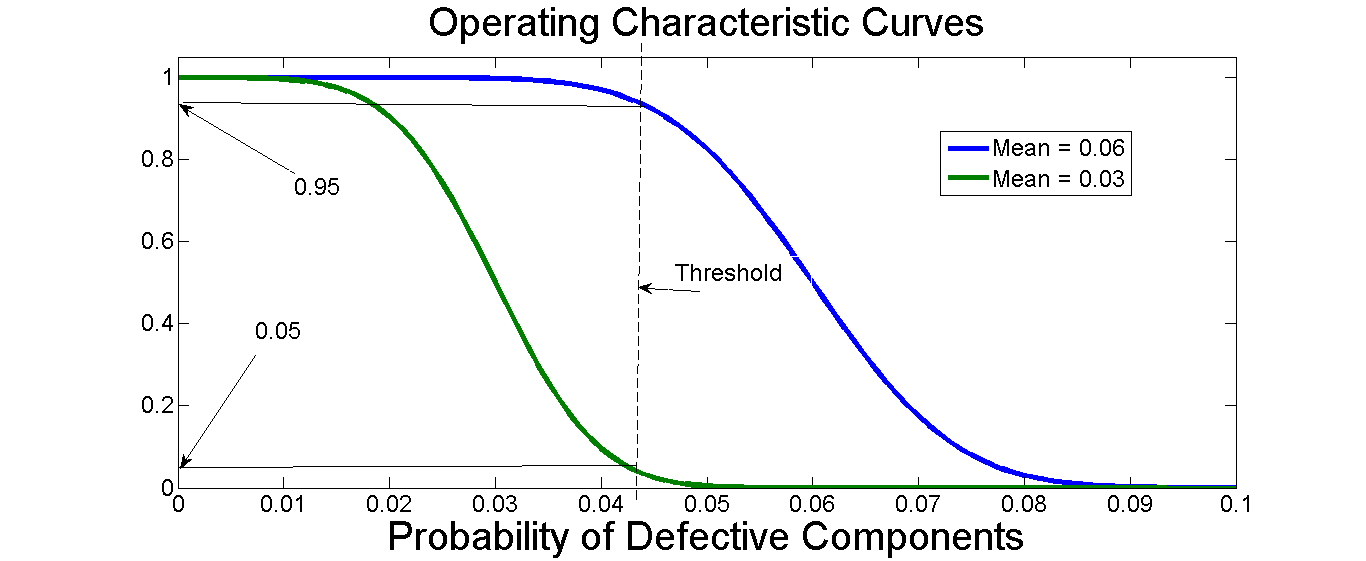}
     \caption{Operating Characteristic - OC curve}
     \label{FigA}
\end{figure*}

 Sequences of success or failure Bernoulli events behave according to the well-known Binomial distribution. In general, the Binomial distribution is not the most suitable (or the easiest) to work with. For small values of $p$ and values of $n \rightarrow \infty$, we can approximate the binomial distribution by the Poisson distribution. We also know that when $np>5$ and $nq>5, q=1-p$ we can use the normal distribution instead. Thus, we have to work with both possibilities: when $np$ and $nq> 5$ and when this is not possible we have to work with the original binomial distribution or with the Poisson approximation. As we work with two distributions side by side (for $p_0$ and $p_1$) and the maximum assumed value for $p_1$ is $<$ 0.5, just test if $n.p_0$, the lesser of the two values, is less than 5. Figure \ref{FigA} illustrates the case with $p_0$ = 0.03 (3\%) and $p_1$ = 0.06 (6\%) considering the same value for the errors $\alpha$ and $\beta$ = 5\%. In this Operation Characteristic Curve, It is possible to see the threshold when Type I and Type II errors are equalized.
 
\subsection{Binomial distribution}
(for \textit{np} and \textit{nq}  $ < 5$) \label{s21}

The cumulative Binomial distribution is $P(X\leq c)=\sum_{i=0}^{i=c}\frac{n!}{(n-i)!}p^i q^{n-i} $ for probability of less or equal $c$ defectives. Thus, the DHT concept is to have two probability distributions, side by side, one with lower mean, $p_0$, and another with higher mean, $p_1$. So, the assumption is that the mean $p_0$ is true (null hypothesis), and the distribution $p_1$ is false (alternative hypothesis). Rejecting $p_0$  is the same as assuming that $p_1$ is the correct distribution for the mean (not rejecting the null hypothesis we are accepting that $p_0$ is true). To make work easier, for small values of $p$ and values of $n \rightarrow \infty$, We can approximate the binomial distribution by the Poisson distribution. As the distribution is discrete in this case (\ref{FigC1}), we have to adjust the values of $n$ and $c$ to allow an acceptable error (depending on the established requirement). In this figure, the distribution on the left will be $Bin(p_0, n)$ and the one on the right will be $Bin(p_1, n)$, representing hypotheses $H_0$ and $H_a$. As can be seen in Fig. \ref{FigC}, for the same value of $n$, the accumulated value reaches 100\% faster with greater $c$ value . By placing the two distributions side by side, it is possible to reconcile the error $1- \frac{\alpha_1}{2}$ of the first ($p_0$) with the value $\frac{\alpha_2}{2}$ of the second ($p_1$). We can notice that the smallest value of $p_0$ (with a given level of significance $\frac{\alpha_1}{2}$ ) whereas the greatest value for $p_1$ (with a given level of significance $\frac{\alpha_2}{2}$). If we are using the same level of significance for both distributions, as $\frac{\alpha_1}{2}$ will match the Type I (producer) error for the distribution $p_0$, the same value corresponds to Type II (consumer) error ($\frac{\alpha_2}{2}$) for the distribution of $p_1$. In situations with different significance levels we establish $ \alpha_1 = \alpha $ and $ \alpha_2 = \beta $.

\subsubsection{Original Binomial distribution}

\begin{figure*}[htb]
     \centering
     \includegraphics[width=0.6\columnwidth]{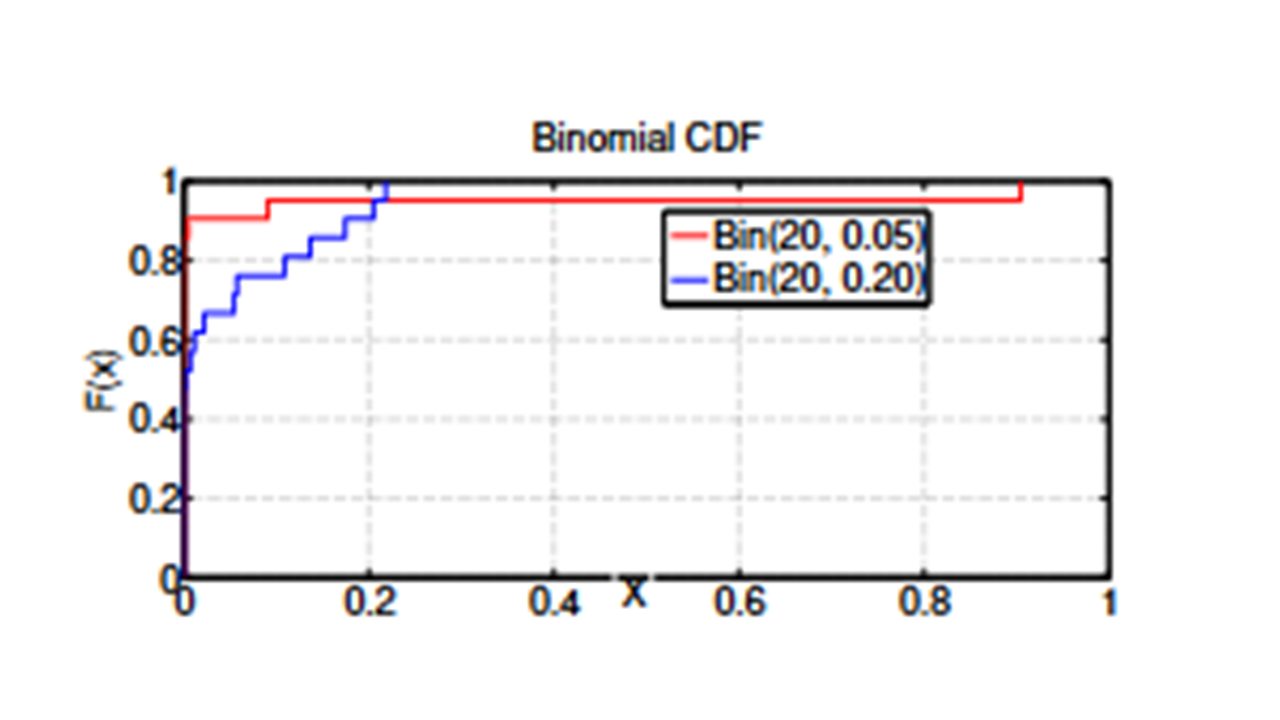}
     \caption{Double Hypothesis Test for Binomial Distribution - Bin Algorithm}
     \label{FigB}
\end{figure*}
The steps to calculate the parameters from Binomial distribution  can be seen in Fig.\ref{FigB} - Bin algorithm. It is necessary to adjust the values of the number of rejections (multiples of $p_0.n$ or $p_1.n$) and round this value to the nearest integer. $L$ = upper limit = $L(p_0)$ and $l$ = lower limit = $l(p_1)$. We calculate the estimated threshold value, $t_h=\frac{L}{n}=\frac{l}{n}$, as the arithmetic mean of these two values, and also the value of $ n.p=n.p_0 $. Variables $x_1$ and $x_2$ are only auxiliary variables. The value $\epsilon_B$ is an admissible error, in most situations we made it initially assumed less than 0.002 (the discrete Binomial distribution did not allows convergence to lower values for the probabilities of interest). For instance, value $\epsilon_B=10^{-4}$ do not allows convergence. We assume $\epsilon_B<0.002$. and  We get $\epsilon_B=0.0019$ to reach the threshold between 0\% and 2\%.  The maximum number of  rejections to accept $H_0$, $c$, will be near $n.t_h$, round($n.t_h$).
\begin{figure*}[htb]
     \centering
   \includegraphics[width=0.5\columnwidth]{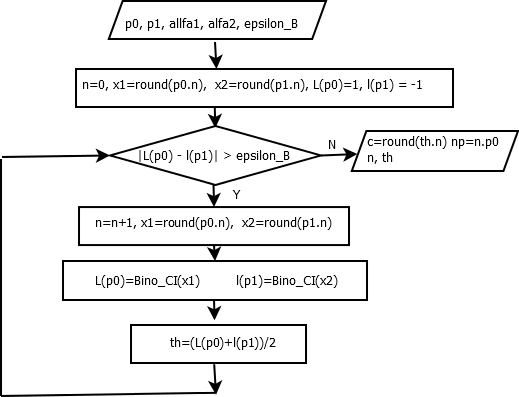}
     \caption{DHT Steps for Binomial Distribution = Bin Algorithm}
     \label{FigC}
\end{figure*}
However, when $p<0.1$ and $n\rightarrow \infty$ we can also apply the Poisson distribution.

\subsubsection{Poisson distribution (for (\textit{np} or \textit{nq}< 5) and \textit{p}<0.1) - Poiss Algorithm}  \label{s22}
\begin{figure*}[htb]
     \centering
     \includegraphics[width=0.5\columnwidth]{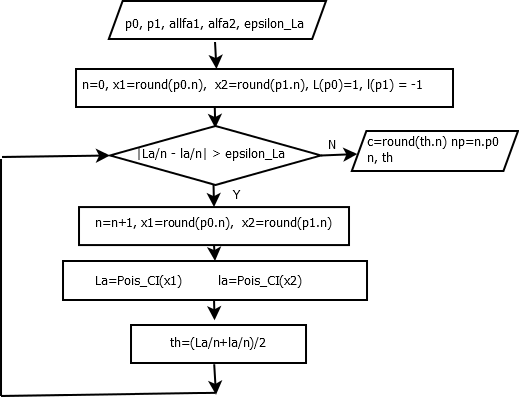}
     \caption{DHT Steps for Poisson Distribution - Poiss Algorithm}
     \label{FigC1}
\end{figure*}

The cumulative distribution approximation for Poisson is $P(X\leq c)=\sum_{i=0}^{i=c}\frac{e^{-\lambda}.{\lambda^i}}{i!} $ (when $p\rightarrow 0$ and $n\rightarrow \infty$, $\lambda=n.p$)  for probability of less or equal $c$ defects.
The steps to calculate the parameters from Poisson distribution  can be seen in Fig.\ref{FigC} - Poiss Algorithm. It is necessary to adjust the values of the number of rejections (multiples of $\lambda=p_0.n$ or $\lambda=p_1.n$) and round this value to the nearest integer. $\frac{\Lambda}{n}$ = upper limit = $\lambda(p_0)$ and $\frac{\lambda}{n}$ = lower limit = $\lambda(p_1)$. We calculate the estimated threshold value, $t_h=(\frac{\Lambda}{n}+\frac{\lambda}{n})/2$, as the arithmetic mean of these two values, and also the value of $ n.p=n.p_0 $. Variables $x_1$ and $x_2$ are only auxiliary variables. The value $\epsilon_{La}$ is an admissible error, in most situations we made it initially assumed less or equal 0.001 (the discrete Poisson distribution did not allows convergence to lower values for the probabilities of interest). For instance, value $\epsilon_{La}=10^{-4}$ do not allows convergence. We assume $\epsilon_{La}\leq0.001$. The maximum number of  rejections to accept $H_0$, $c$, will be near $n.t_h$, round($n.t_h$). The steps for calculate the Poisson distribution can be seen in \ref{FigC1}. 

However, when the values of $ n.p> 5 $, we can use the Normal distribution, which has the facility of being a continuous distribution.

\subsection{Normal distribution (for \textit{np} and \textit{nq}> 5) \label{s23}}

We know that the sample distribution of any kind of statistical distribution converges, by the Central Limit Theorem, to the Normal distribution, $$N\sim \big(\overline{x},\dfrac{\sigma^2}{n}\big) $$ where $\overline{x}$  is the sample mean and $\sigma_{\overline{x}}^2=\dfrac{\sigma^2}{n}$ is the sample variance. In Bernoulli distribution, $\overline{x}=\hat{p}$, the average probability of success in each experiment and  $\dfrac{\sigma}{\sqrt{n}}=\dfrac{\sqrt{\hat{p}(1-\hat{p})}}{\sqrt{n}}=\sqrt{\dfrac{\hat{p}(1-\hat{p})}{n}}$ is the sample standard deviation, where $\hat{p}(1-\hat{p})$ is its variance. 

Illustration of DHT for normal distribution with the probabilities side by side can be seen in Fig. \ref{FigF} and  in Fig. \ref{FigG}. We can write the boundaries of the two distributions as follows ($\alpha_1 = \alpha_2= \alpha$):
\begin{equation} \label{eq1}
    \hat{p_0}+z_{\frac{\alpha}{2}}\sqrt{\dfrac{\hat{p_0}(1-\hat{p_0})}{n}}
\end{equation}
\begin{equation} \label{eq2}
    \hat{p_1}-z_{\frac{\alpha}{2}}\sqrt{\dfrac{\hat{p_1}(1-\hat{p_1})}{n}}
\end{equation}

Equation (\ref{eq1}) represents the upper limit for and equation (\ref{eq2}) represents the lower limit for. In practice we will perform $n$ Bernoulli experiments in which we will have a certain number of failures (and successes). Let's assume that in the case of distribution in (\ref{eq1})  we have $L$ failures and in distribution (\ref{eq2})  we have $l$ failures (in both cases, $n$ attempts). Then it is possible to rewrite (\ref{eq1}) and (\ref{eq2}):
\begin{equation} \label{eq3}
   \dfrac{L}{n}= \hat{p_0}+z_{\frac{\alpha}{2}}\sqrt{\dfrac{\hat{p_0}(1-\hat{p_0})}{n}}
\end{equation}
\begin{equation} \label{eq4}
   \dfrac{l}{n}= \hat{p_1}-z_{\frac{\alpha}{2}}\sqrt{\dfrac{\hat{p_1}(1-\hat{p_1})}{n}}
\end{equation} 
 
Another way to verify the system equations is to multiply everything by $n$. The result is that the Bernoulli event number values, $L$ and $l$, as expected, are distributed according to a Binomial distribution of mean $np$ and variance $np(1-p)$:
 
We can notice that equation (\ref{eq3}) corresponds to the largest value of $p_0$ (with a given level of significance $\frac{\alpha_1}{2}$) whereas equation (\ref{eq4}) corresponds to the smallest value for $p_1$ (with a given level of significance $\frac{\alpha_2}{2}$). As $\frac{\alpha_1}{2}$ will match the Type I (producer) error for the distribution $p_0$, $\frac{\alpha_2}{2}$ will corresponds to Type II (consumer) error for the distribution of $p_1$. Therefore, solving the systems with equations (\ref{eq3}) and (\ref{eq4}) will define the number of tests and rejections ($n, c$) to see whether or not the null hypothesis will be rejected. As this is also a system of nonlinear equations, its exact solution can be done by Newton Raphson's generalized method (\ref{swc3}). 
\subsubsection{Resolution of the Nonlinear System by Generalized Newton-Raphson Method - Norm\_N Algorithm}
\label{swc3}

The system formed by equations (\ref{eq3} and \ref{eq4}) can be rewritten as:
 
\begin{equation} \label{eq5}
  F_1= \dfrac{L}{n}- \hat{p_0}-z_{\frac{\alpha}{2}}\sqrt{\dfrac{\hat{p_0}(1-\hat{p_0})}{n}}
\end{equation}
\begin{equation} \label{eq6}
  F_2= \dfrac{l}{n}- \hat{p_1}+z_{\frac{\alpha}{2}}\sqrt{\dfrac{\hat{p_1}(1-\hat{p_1})}{n}}
\end{equation} 
Thus, the previous equations will be both set equal to zero and call them $F_1$ and $F_2$.
We assume that $\frac{L}{n}=\frac{l}{n}$   so that both errors are equal. That way, we will do $x_1=\frac{L}{n}=\frac{l}{n}$ and $x_2=n$. 
The equations now are:
   $X= 
  \begin{bmatrix}
    x_1\\
    x_2
  \end{bmatrix} $
    and
    $F= 
  \begin{bmatrix}
    F_1\\
    F_2
   \end{bmatrix}$,
    where:
    \begin{equation} \label{eq7}
  F_1= x_1- \hat{p_0}-z_{\frac{\alpha}{2}}\sqrt{\dfrac{\hat{p_0}(1-\hat{p_0})}{x_2}}
\end{equation}
\begin{equation} \label{eq8}
  F_2= x_1- \hat{p_1}+z_{\frac{\alpha}{2}}\sqrt{\dfrac{\hat{p_1}(1-\hat{p_1})}{x_2}}
\end{equation} 
 Jacobian is formed by partial derivatives:\\
\begin{center}
\begin{equation}
JF=
\begin{bmatrix} 
\dfrac{\partial{F_1}}{\partial{x_1}}&\dfrac{\partial{F_1}}{\partial{x_2}}\\
\dfrac{\partial{F_2}}{\partial{x_1}}&\dfrac{\partial{F_2}}{\partial{x_2}}
 \end{bmatrix}=
\begin{bmatrix} \label{eq9}
1 \hbox{  }\dfrac{z_\frac{\alpha}{2}}{2\sqrt{p_0(1-p_0)x_2^3}}\\
1 \hbox{  } \dfrac{-z_\frac{\beta}{2}}{2\sqrt{p_1(1-p_1)x_2^3}}\\
 \end{bmatrix}
 \end{equation}
\end{center}

\begin{figure*}[htb]
     \centering
     \includegraphics[width=1\columnwidth]{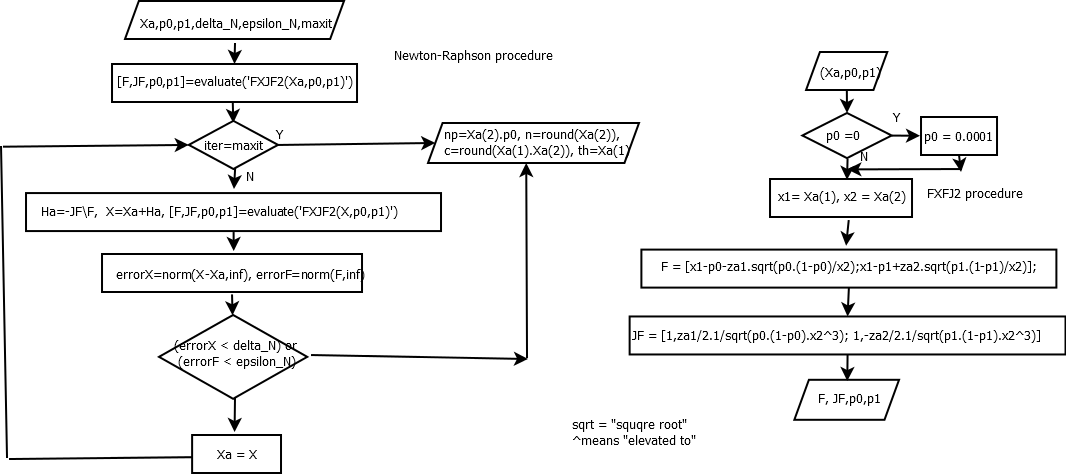}
     \caption{DHT for Normal Newton Raphson Generalized Algorithm - Norm\_N}
     \label{FigD}
\end{figure*}

Newton-Raphson's generalized matrix method resolution can be seen in Ng and Lee (2000) and Skaflestad (2003), and the program can be obtained from the Faculty Washington (2017) web site https://online.stat.psu.edu/stat414/node/179/. Its use requires little adaptation to equations (\ref{eq9}) and its Matlab program as in Fig. \ref{FigD}.
Equations (\ref{eq1}...\ref{eq9}) are then solved by the Newton-Raphson method. The algorithm has three stopping possibilities: a) by reaching the maximum number of iterations ($maxit$), b) by the Euclidean norm of the vector $X$ (whose variables are $X(1)$ - the threshold sought, and $X(2)$ - the number of tests to be carried out), have a value less than $\delta_N=10^{-9}$, or c) by the Euclidean norm of the vector $F$ (to be zeroed), be less than $\epsilon_N=10^{-8}$. On the other hand, this method also allows to have a certain control of the number of iterations, that is, knowing that the input vector $X_a$ has $X_a(1)$ related to the threshold and $X_a(2)$ to the number of Bernoulli trials ($ n $).
\subsubsection{More Intuitive Solution for Normal Distribution - Norm\_I Algorithm \label{sub1}}
\begin{figure*}[htb]
     \centering
     \includegraphics[width=0.5\columnwidth]{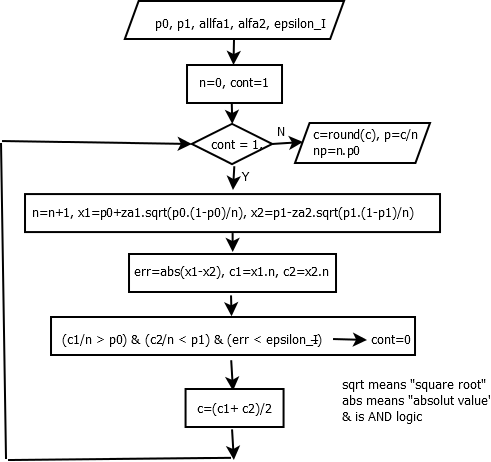}
     \caption{DHT for Intuitive Normal Algorithm -Norm\_I}
     \label{FigE}
\end{figure*}

However, because the number of units is integer, there is a simple way to solve the system by adding one unit at a time and evaluating the error, as in Fig.\ref{FigE} - Norm\_I Algorithm. As we know that the number of tests is an integer value, $ n $, we start with $ n = 1 $ and increase its value to observe the convergence when the upper limit and the lower are equal. The same Equations (\ref{eq3} and \ref{eq4}) are used. The stopping criterion corresponds to a proposed error, $\epsilon_I=10^{-4}$ or $\epsilon_I=10^{-6}$, which indicates when the limit values are close enough. The two algorithms for the Normal distribution return the number of tests, $n$, the maximum number of rejections, $c$, the number that allows Binomial to Normal approximation, $n.p$, and the threshold $p$.
\section{Examples for the Four Approaches}

\begin{figure*}[htb]
     \centering
     \includegraphics[width=0.8\columnwidth]{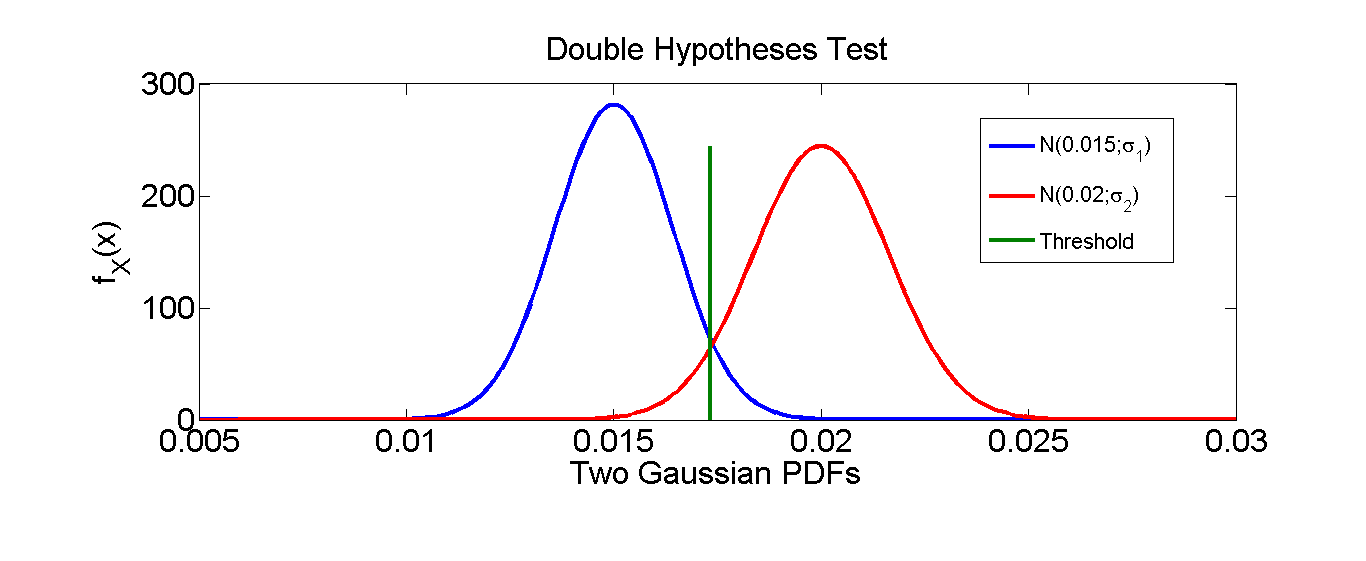}
     \caption{DHT for discrimination between two Gaussians}
     \label{FigF}
\end{figure*}

Considering initially the situation in which we want to decide between two very close values, e.g. $p_o$ = 0.015 (or 1.5\%) and $p_1$  = 0.02 (or 2.0\%).  The idea of DHT is to have a control of both types of errors (Type I and Type II, or $\alpha$ and $\beta$. or producer and consumer). Assuming both confidence levels will be 5\%, we do $z_{\frac{\alpha}{2}} = 1.64$. Fig.\ref{FigF} shows the two Gaussian PDFs where you can see that choosing $p_0$ implies in 5\% the probability of really be $p_1$, and choosing $p_1$ implies in 5\% the probability of really be $p_0$.
Initially, considering Newton-Raphson solution, the value obtained for  $$X= 
  \begin{bmatrix}
    \dfrac{L}{n}=\dfrac{l}{n}\\
    n
  \end{bmatrix} 
  =
   \begin{bmatrix}
    0.0173\\
    7359.8
  \end{bmatrix}$$
The value 0.0173 is the threshold between the two distributions that appears on Fig. \ref{FigF}. The high value found for $n$ is because the discrimination occurs for values very close. Doing $L = l$ = 0.0173 $\times$ 7359.8 = 128 rounded up. Thus, $L = l=c$ = 128 and $n = 7360$ (with rounding) because these values should be integers (using simplest recurring procedure for Normal distribution with $\epsilon_I=10^{-6}$, as in Subsection \ref{sub1}, we get $n$ = 7357 and threshold = 0.0173, the same value). Standard deviations are $\sigma_1$ = 0.0014 and $\sigma_2$  = 0.0016. Interpretation of the result is as follows: if in 7360 Bernoulli repetitions occur less than 128 “failures” we can say that the distribution is at least that of  (may even be better than $p_0$, that is, have less probability). If 128 or more “failures”, occur we can say that the distribution is that of $p_1$  (may be worse than $p_1$ if more failures occur than $p_1$). So, when you reject $p_0$, If you want to know the percentage of “failures”, you can continue the test to another level, say $p_2$ (provided that the requirement allows higher probabilities). The next test will show a
situation where three “failure” levels are defined: 2\%, 5\% and 10\%, for instance. Tests are always performed two by two.
Initially compares 2\% with 5\% and, if 2\% was rejected, then compares  5\% with 10\%. Tables \ref{Tab10}, \ref{Tab101}, \ref{Tab1} and \ref{Tab11} show the results obtained for the significance level of 5\% $(z_{\frac{\alpha}{2}}=z_{\frac{\beta}{2}}=1.64)$ for the four procedures. Tables \ref{Tab10} and \ref{Tab101}, with Bin and Poiss procedure, are the only that can calculate approached values starting from zero ($n.p =n.p_0<5 $).
\begin{center}
\begin{table}[h]
\centering
\caption{{\bf Tests using {\bf Bin} }
{\bf with $\epsilon_B=0.0019$}}
\label{Tab10}
\begin{tabular}{|c|c|c|}
\hline
{\bf Test 0 Bin}&{\bf Test 1 Bin} & {\bf Test 2 Bin}  \\
\hline
0\% and 2\% & 2\% and 5\% & 5\% and 10\% \\
\hline
$t_h=0.0095$ & $x_1=t_h$ = 0.0348 & $t_h$ = 0.0744 \\
\hline
$n=375$ & $n=550$ & $n=405$\\
\hline
$n.p=n.p_0 = 0$ & $n.p=n.p_0 = 0.02\times 550 =11$ & $n.p=n.p_0 = 0.05\times 405=20.25$ \\
\hline
$L = l =c =4$ &$L=l=c =19$ & $ L = l =c=30 $ \\
\hline
$\sigma_1$=0 & $\sigma_1$=0.0072 & $\sigma_1$=0.0093 \\
$\sigma_2$=0.0060& $\sigma_2$=0.0108& $\sigma_2$=0.0149\\
\hline
\end{tabular}
\end{table}
\end{center}
\begin{center}
\begin{table}[h]
\centering
\caption{{\bf Tests using {\bf Poiss} }
{\bf with $\epsilon_{La}=0.001$}}
\label{Tab101}
\begin{tabular}{|c|c|c|}
\hline
{\bf Test 0 Poiss}&{\bf Test 1 Poiss} & {\bf Test 2 Poiss}  \\
\hline
0\% and 2\% & 2\% and 5\% & 5\% and 10\% \\
\hline
$t_h=0.0095$ & $x_1=t_h$ = 0.0343 & $t_h$ = 0.0742 \\
\hline
$n=375$ & $n=570$ & $n=465$\\
\hline
$n.p=n.p_0 = 0$ & $n.p=n.p_0 = 0.02\times 570 =11.4$ & $n.p=n.p_0 = 0.05\times 465=23.25$ \\
\hline
$L = l =c =4$ &$L=l=c =20$ & $ L = l =c=35 $ \\
\hline
$\sigma_1$=0 & $\sigma_1$=0.0072 & $\sigma_1$=0.0093 \\
$\sigma_2$=0.0060& $\sigma_2$=0.0108& $\sigma_2$=0.0149\\
\hline
\end{tabular}
\end{table}
\end{center}

\begin{center}
\begin{table}[h]
\centering
\caption{{\bf Values for three-level “failure” testing using {\bf Norm\_N}}
 }
\label{Tab1}
\begin{tabular}{|c|c|}
\hline
{\bf Test 1 Norm\_N} & {\bf Test 2 Norm\_N}  \\
\hline
2\% and 5\% & 5\% and 10\% \\
\hline
 $x_1=t_h$ = 0.0317 & $x_1=t_h$ = 0.0710 \\
\hline
 $x_2 = 382.89 \rightarrow n=383$ & $x_2 = 288.61\rightarrow n=289$\\
\hline
  $n.p=n.p_0 = 0.02\times 383=7.66$ & $n.p=n.p_0 = 0.05\times 289=14.45$ \\
\hline$ L = l =c =\lceil{x_1\times x_2}\rceil=13$ & $ L = l =c= \lceil{x_1\times x_2}\rceil=21 $ \\
\hline 
$\sigma_1$=0.0072 & $\sigma_1$=0.0128(*) \\
 $\sigma_2$=0.0111& $\sigma_2$=0.0176\\
\hline
 \multicolumn{2}{|c|}{(*) This value was used in both graphs: as $\sigma_2$ in Test 1 and as $\sigma_1$ in Test 2, }\\
 \multicolumn{2}{|c|}{where $\lceil{x}\rceil$ is the smallest integer greater than $x$.  }\\
\hline
\end{tabular}
\end{table}
\end{center}

\begin{figure*}[htb]
     \centering
     \includegraphics[width=0.8\columnwidth]{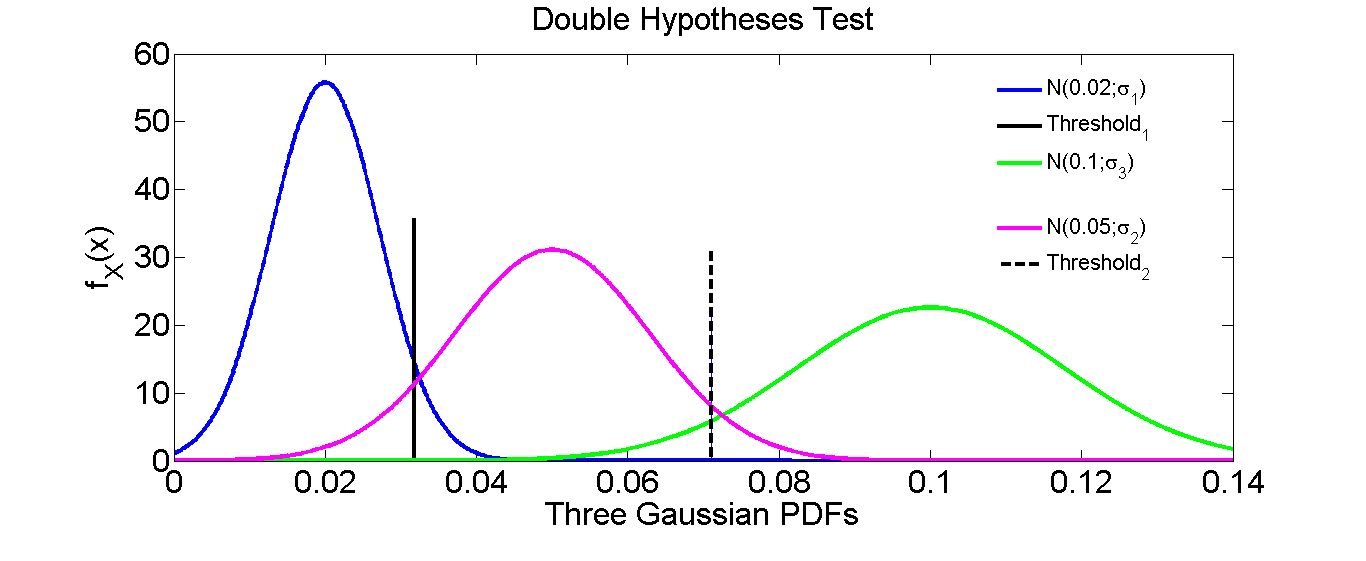}
     \caption{DHT for discriminaton between three Gaussians}
     \label{FigG}
\end{figure*}

On the other hand, Table \ref {Tab11} shows the same calculation made for the most intuitive method, Subsection \ref{sub1}, corresponding to columns two and three (Tests 1 and 2). Column with Test 0 can only be filled in using the Binomial or Poisson distributions method (Subection \ref{s21}) because it always starts with 0\% of failures ($ n.p_0 $ = 0).
\begin{center}
\begin{table}[h]
\centering
\caption{{\bf Tests using  {\bf Norm\_I}}
 {\bf with $\epsilon_I=10^{-4}$}}
\label{Tab11}
\begin{tabular}{|c|c|}
\hline
{\bf Test 1 Norm\_I} & {\bf Test 2 Norm\_I}  \\
\hline
 2\% and 5\% & 5\% and 10\% \\
\hline
 $x_1=t_h$ = 0.0315 & $t_h$ = 0.0694 \\
\hline
 $n=381$ & $n=288$\\
\hline
 $n.p=n.p_0 = 0.02\times 381=7.62$ & $n.p=n.p_0 = 0.05\times 288=14.40$ \\
\hline
$L=l=c =12$ & $ L = l =c=20 $ \\
\hline 
 $\sigma_1$=0.0072 & $\sigma_1$=0.0128 \\
 $\sigma_2$=0.0112& $\sigma_2$=0.0177\\
\hline
\end{tabular}
\end{table}
\end{center}
Fig. \ref{FigG} shows three curves with the Gaussian PDFs (the values of $x_1$ are the values of the thresholds between the curves). The three curves were plotted, but the variance of the second curve ($\sigma_2$ from the first test was used) is slightly different from the real one because the number of attempts ($n$) for the second test is different from the first one (there should be two sets of two curves, one set for each test, but the difference is minimal and the illustration would look similar). From a didactic point of view it is important to show the three curves together. In this case, the interpretation is as follows (valid for Norm\_N but similar for Norm\_I, Poiss and Bin). If there are less than 13 “failures” in 383 attempts ($L, l < 13$ ),the level is better than or equal to 2\%. If there are  $13 \leq L, l < 21$  “failures” in 289 attempts, the level is 5\%. If there are more than 21 “failures” in 289 attempts ($L, l > 21$),the level is less than or equal to 10\%. For instance, if we want to evaluate the next level we need to compare 10\% with 15\%, for instance. It can be noted that, whenever possible, it pays to use continuous distributions, except when you want to get close to zero. Often in practice it is not necessary to include zero (0 \%) in the assessment. For example, you can start with 0.5 \%. In the Section 5 more extreme examples. 
\section{Successive Failures Limit (SFL)}
According to Feller (1968), the “time” $E[X]$ (we observe that time is computed by the number of occurrences) of recurrence (so that a given sequence of “$r$” failures recurs), computed in number of Bernoulli’s occurrences. is:
       \begin{equation} \label{eqsfl}
           E[X]=\dfrac{1-p^r}{log(p)}
       \end{equation}
    
where $p$ is the probability of "success" and $q = 1- p$ is the probability of "failure" of each individual Bernoulli event. Thus, by inverting equation (\ref{eqsfl}) using the contraction theorem around a fixed point, Istratescu (1981),
\begin{equation} \label{eqsfl1}
           r=\dfrac{log\Big(\dfrac{1-p^r}{E[X]q}\Big)}{log(p)}
       \end{equation}

\begin{figure*}[htb]
     \centering
     \includegraphics[width=0.6\columnwidth]{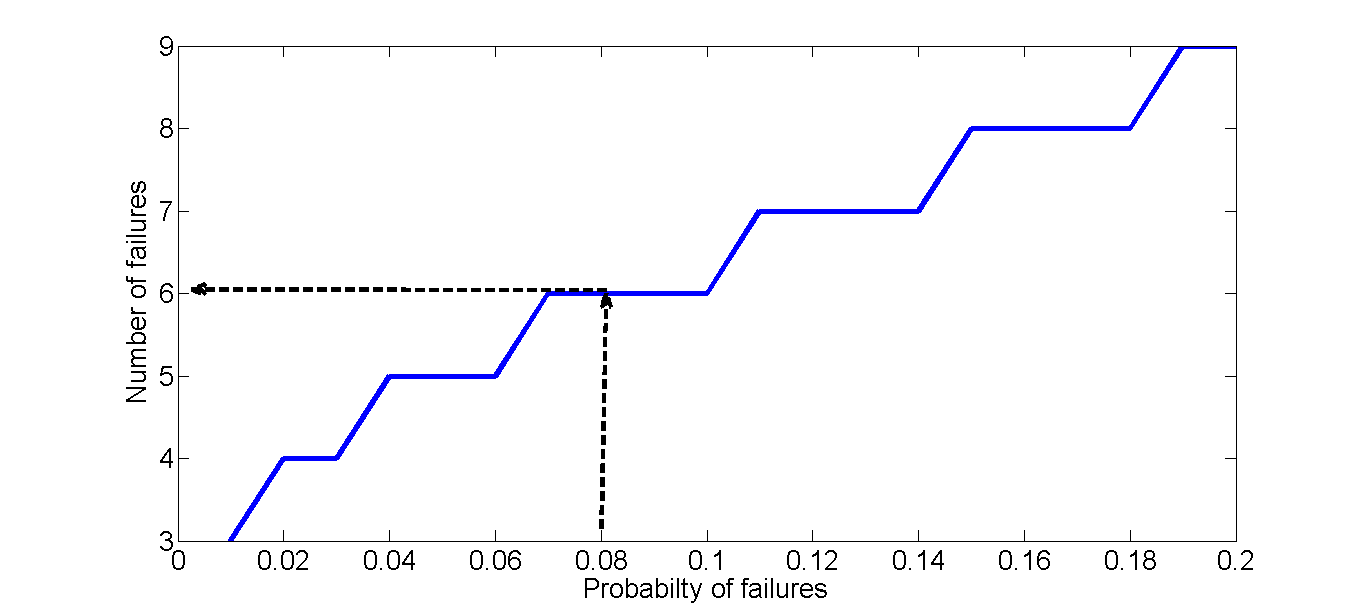}
     \caption{Successive Failures Limit}
     \label{FigH}
\end{figure*}

We can observe that the variable $r$ on the left side is a function of $r$ on the right side ($r = f(r)$). For example, for 
$E[X] = 10^6$ Bernoulli events and $p = 0.02$, initial value of $r, r_0 = 1$, $r=4$  is obtained (as in Fig. \ref{FigH}). The interpretation is that every $10^6$ events occur a single sequence with $r = 4$ “failures” on average. So, for more than four successive failures the level must be rejected. For $p = 0.05, r = 5$, and for $p = 0.10, r = 6$. The application of the value $r$, the successive failures limit, must be done in conjunction with the DHT. In the case previously seen, for 1, 2, 5 and 10\%, for example, with $E[X] = 10^6$, the interpretation is as follows: If a sequence of 
$r = 3$ successive “failures” occurs, it can be concluded that the percentage of failures is greater than 1\%. If a sequence of $r = 4$ “failures” occurs, it can be concluded that the percentage of failures is greater than 2\%, and if $r = 6$ occurs, it can be concluded that the percentage is greater than 10\%.
\section{More Extreme Examples, Discussion and Considerations}
Table \ref{Tab0} shows several situations in which the four methods are applied. In this table two more parameters were added (last two columns), the execution time, $ t_{exec} $ obtained for the same hardware using Matlab's tic and toc functions, and the number of iterations, $ iter $. Paraneter $ iter $ is different from the $ n $ value (Bernoulli trials) only for Norm \_N method because it uses its own iterative method with derivatives. The  $t_{exec} $ field corresponds to the time elapsed in the same  hardware.

\begin{center}
\begin{table}[hp]
\centering
\caption{{\bf Bernoulli trials, rejection number and threshold}
 }
\label{Tab0}
\begin{tabular}{|c|c|c|c|c|c|c|c|}
\hline
\multicolumn{8}{|c|}{{\bf Some Values for probability steps}}\\
\hline \hline
Method & interval & $n$ &   $L = l=c$ &  $t_h$ &  $n.p$ &   $t_{exec}$ &  $iter$ \\
\hline \hline
Bin     &  0--0.01 & 650 &3    & 0.0050 & 0 &  0.2301 & 650 \\
\hline
Poiss     &  0--0.01 & 750 &4    & 0.0048 & 0 &  1.8657 & 750 \\
\hline

Norm\_{N} (**)     &  0--0.01 & 328  & 0    & 0.0010 & 0.0328 &  0.8946 & 10000 \\
\hline
Norm\_{I} (*)    &  0--0.01 & 267 & 0    & 0 & 0 &  0.00004 & 267 \\
\hline
\hline
Bin     &  0.01--0.02 & 1645 & 24    & 0.0148 & 16.45 &  0.7088 & 1645 \\
\hline
Poiss    &  0.01--0.02 & 1939 & 29    & 0.0148 & 19.39 &  2.2218 & 1939 \\
\hline
Norm\_{N}     &  0.01--0.02 & 1543 & 22    & 0.0142 & 15.43 &  0.1875 & 1238 \\
\hline
Norm\_{I}     &  0.01--0.02 & 1543 & 22    & 0.0143 & 15.43 &  0.0003 & 1543 \\
\hline
\hline
Bin     &  0.02--0.03 & 2717 & 68    & 0.0250 & 54.34 &  1.0446 & 2717 \\
\hline
Poiss     &  0.02--0.03 & 3217 & 80    & 0.0249 & 64.34 &  3.8927 & 3217 \\
\hline

Norm\_{N}     &  0.02--0.03 & 2594 & 64    & 0.0245 & 51.89 &  0.2284 & 730 \\
\hline
Norm\_{I}     &  0.02--0.03 & 2594 & 64    & 0.0247 & 51.88 &  0.0002 & 2594 \\
\hline \hline
Bin     &  0.07--0.08 & 7607 & 570    & 0.0750 & 532.49 &  2.8932 & 7607\\
\hline
Poiss     &  0.07--0.08 & 9407 & 704    & 0.0748 & 658.49 &  11.657 & 9407\\
\hline

Norm\_{N}     &  0.07--0.08 & 7454 & 558    & 0.0748 & 521.77 &  0.0271 & 258 \\
\hline
Norm\_{I}     &  0.07--0.08 & 7453 & 558    & 0.0749 & 521.71 &  0.0006 & 7453 \\
\hline \hline
Bin     &  0--0.2 & {\bf N/A} & {\bf N/A}    & {\bf N/A} & {\bf N/A} &  {\bf N/A} & {\bf N/A} \\
\hline
Poiss     &  0--0.2 & {\bf N/A} & {\bf N/A}    & {\bf N/A} & {\bf N/A} &  {\bf N/A} & {\bf N/A} \\
\hline

Norm\_{N} (*)     &  0--0.2 & 11 & 0    & 0.0050 & 0.0011 &  0.4700 & 4522 \\
\hline
Norm\_{I}     &  0--0.2 & {\bf N/A} & {\bf N/A}    & {\bf N/A} & {\bf N/A} &  {\bf N/A} & {\bf N/A} \\
\hline \hline
Bin     &  0.2--0.4 & 82 & 24    & 0.2965 & 16.40 &  0.0338 & 82 \\
\hline
Poiss    &  0.2--0.4 & 117 & 35    & 0.2951 & 23.40 &  0.1707 & 117 \\
\hline

Norm\_{N}     &  0.2--0.4 & 53 & 15    & 0.2899 & 10.6497 &  0.0098 & 89 \\
\hline
Norm\_{I}     &  0.2--0.4 & {\bf N/A} & {\bf N/A}    & {\bf N/A} & {\bf N/A} &  {\bf N/A} & {\bf N/A} \\
\hline
\hline
 \multicolumn{8}{|c|}{(*) These values are meaningless because $ n.p <5$. For Norm\_I $\epsilon_I = 10^{-6}$.}\\
 \multicolumn{8}{|c|}{(**) $ n.p <5$ and  the max. \# of iter. for Norm\_N, 10000, has been reached.}\\
  \multicolumn{8}{|c|}{Used  $\epsilon_B=0.0019$ and $\epsilon_{La}=0.001$. {\bf           N/A = not available}. .}\\
   \multicolumn{8}{|c|}{Precision for Norm \_N is $10^{-8} $ or $10^{-9}$}\\
\hline

\end{tabular}
\end{table}
\end{center}

In Table \ref{Tab0}, first  two lines, it can be seen that the application of the Bin or Poiss method is the only possible one because the others have $ n.p_0 = 0 $. However, the Bin method is not recommended when $ n.p_0> 5 $ as it presents higher values of Bernoulli's trials compared to the others, which is not appropriate. The best and most accurate method is Norm \_N, which warns even when $ n.p <5 $. In most practical situations, the simplest to program Norm \_I can be applied, or with $ \epsilon_I = 10 ^ {- 4} $, or with $ \epsilon_I = 10 ^ {- 6} $, depending on the required accuracy . However, Norm \_I does not converges in some situations, as can be seen in the last lines of Table \ref{Tab0}. It should also be noted the issue of execution time, $t_{exec}$, which can be significantly higher in the case of Norm\_N compared to Norm\_I. This could be very important in real time applications (for instance, comparing between 0.01 and 0.02, as in Table \ref{Tab0}, Norm\_I performs 625 times faster than Norm\_N). However, as Norm\_N uses derivatives, it also has a smaller number of iterations, $iter$, to converge. This can, in different situations, mean less processing time.
\section{Fuzzy Design}

\begin{figure*}[htb]
     \centering
     \includegraphics[width=0.98\columnwidth]{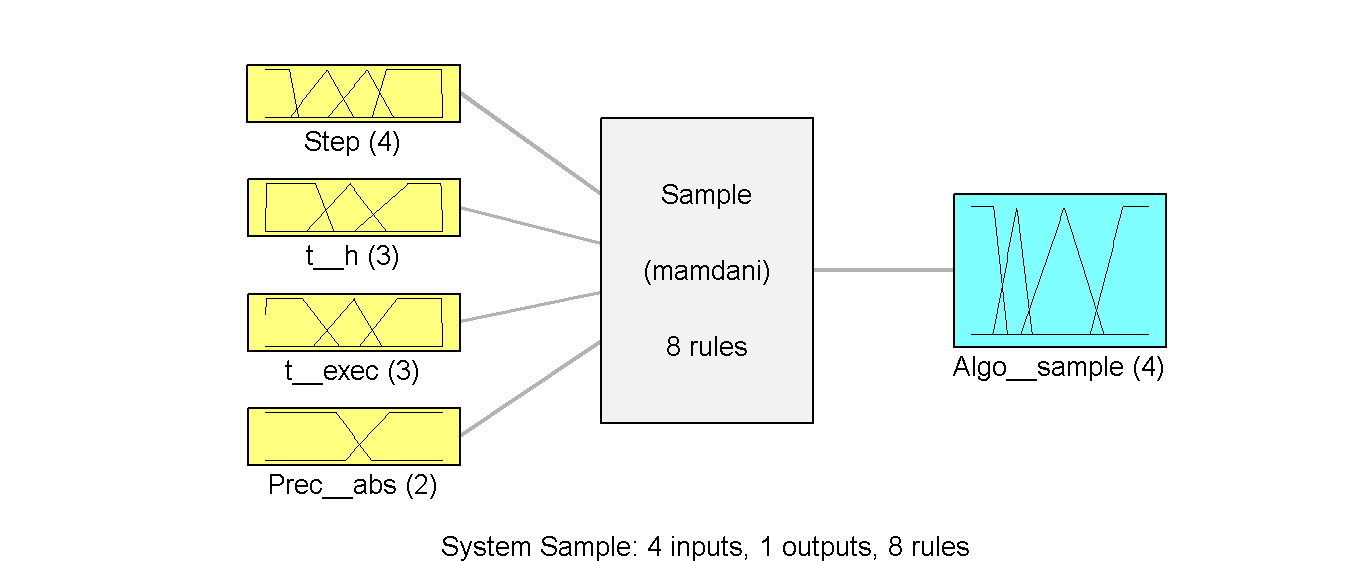}
     \caption{Fuzzy Design}
     \label{FigI}
\end{figure*}

\begin{figure*}[htb]
     \centering
     \includegraphics[width=0.98\columnwidth]{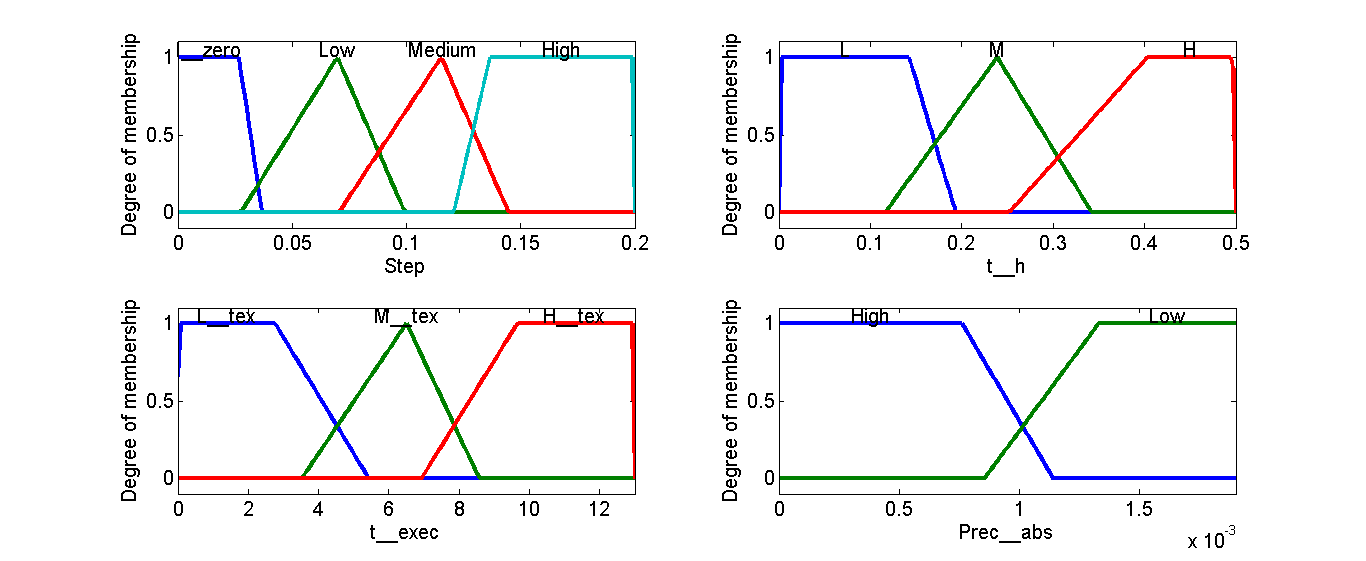}
     \caption{Fuzzy Inputs}
     \label{FigJ}
\end{figure*}

\begin{figure*}[hp]
     \centering
     \includegraphics[width=0.98\columnwidth]{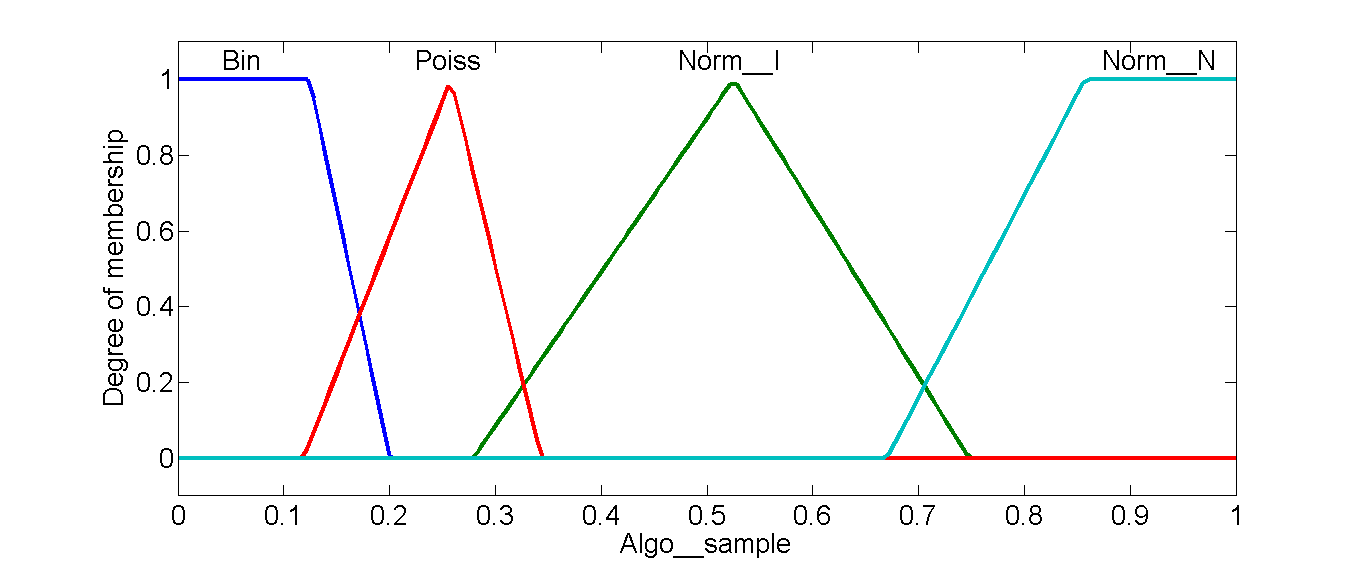}
     \caption{Fuzzy Output}
     \label{FigK}
\end{figure*}

\begin{figure*}[hp]
     \centering
     \includegraphics[width=0.98\columnwidth]{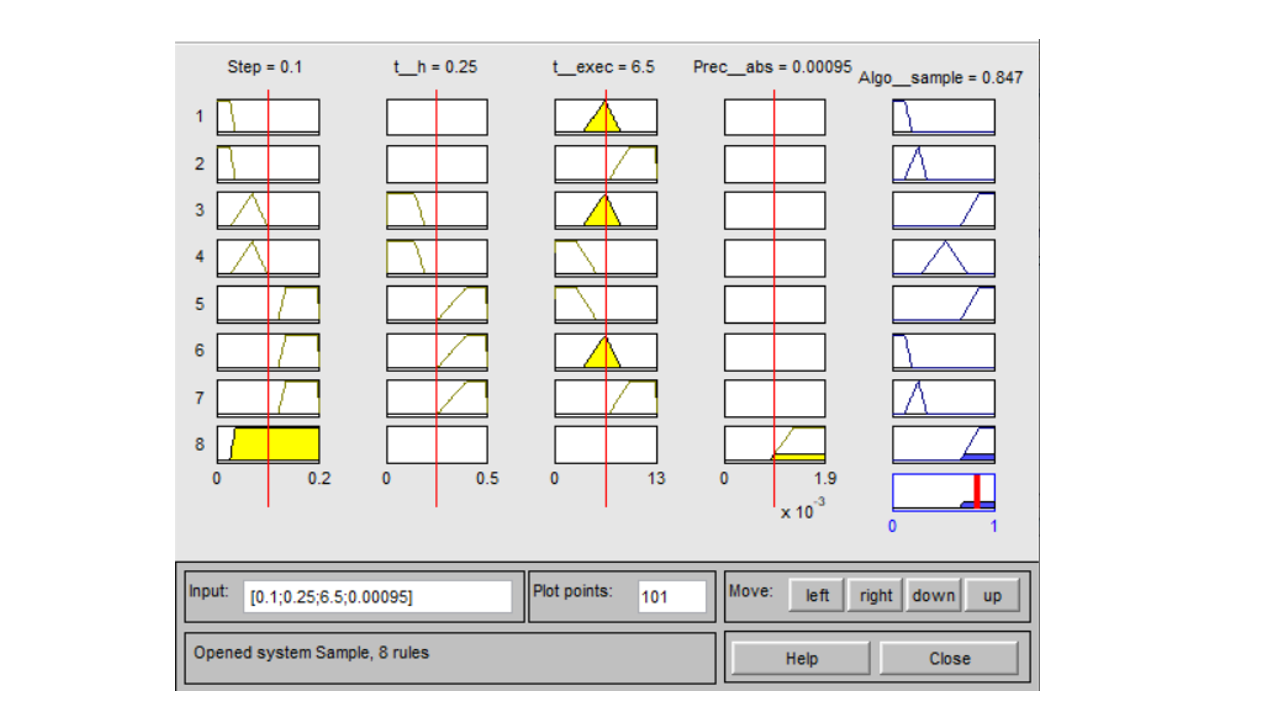}
     \caption{Fuzzy Rules View}
     \label{FigL}
\end{figure*}

Four main variables should guide the user's decision making: a) $Step$ - the interval (precision) to be obtained for the estimated value $\hat{p}$, b) $ t_h $ - the threshold between two possible estimates for $\hat{p}$, c) $ t_ {exec} $ in relative units, obtained for the same hardware, and d) $Prec_ {abs}$, the absolute precision for obtaining the threshold value $ t_h $. The output will be the indication of the best algorithm for each set of inputs. The Fuzzy procedure uses the Mamdani method, with four inputs, one output and eight rules, see Fig\ref{FigI}.
With respect to the four input variables, we can say that: the input $Step$ has a range of 0 to 0.2 (0 to 20.0 \%). The variable $t_h$ can vary between 0 to 0.5 (0 to 50.0 \%) since it must be placed between two possible values of $\hat{p}$. The variable $t_{exec}$ is essential when the algorithm is applied in real time. It has a variation between 0 and 12 units, whose value reflects the relative execution time of the algorithms on the same hardware. The variable $Prec_{abs}$ can vary between 0 and 10$^{-3}$, that is, the lower its value, the greater the precision. Figure \ref{FigJ} shows the four input variables.
The output variable, $Algo\_sample$, shows the defuzzification value, ranging from 0.0 to 1.0, and providing an indication of the most appropriate algorithm in view of the values of the input variables. The association of the algorithms with the output value is: Bin from 0.1 to 0.15, Pois from 0.15 to 0.32, Norm\_I from 0.32 to 0.71 and Norm\_N for values greater than 0.71. Figure \ref{FigK} shows the output variable $Algo\_sample$.
The following eight Fuzzy rules are used to calculate the defuzzification value using the centroid method:
\begin{enumerate}
\item If ($Step$ is $I\_zero$) and ($t_{exec}$ is $M\_tex$) then ($Algo\_sample$ is Bin)                
\item If (Step is $I\_zero$) and ($t_{exec}$ is $H\_tex$) then ($Algo\_sample$ is Poiss)                
\item If ($Step$ is Low) and ($t_h$ is L) and ($t_{exec}$ is $M\_tex$) then ($Algo\_sample$ is Norm\_N)
\item If ($Step$ is Low) and ($t_h$ is L) and ($t_{exec}$ is $L\_tex$) then ($Algo\_sample$ is Norm\_I)
\item If ($Step$ is High) and ($t_h$ is H) and ($t_{exec}$ is $L\_tex$) then ($Algo\_sample$ is Norm\_N) 
\item If ($Step$ is High) and ($t_h$ is H) and ($t_{exec}$ is $M\_tex$) then ($Algo\_sample$ is Bin)     
\item If ($Step$ is High) and ($t_h$ is H) and ($t_{exec}$ is $H\_tex$) then ($Algo\_sample$ is Poiss)  
\item If ($Step$ is not $I\_zero$) and ($Prec\_abs$ is Low) then ($Algo\_sample$ is Norm\_N)         
\end{enumerate}
An example of defuzzification can be seen in Fig. \ref{FigL} and the indication of a value greater than 0.8, suggesting the use of the Norm\_N algorithm.

\begin{figure*}[hp]
     \centering
     \includegraphics[width=0.98\columnwidth]{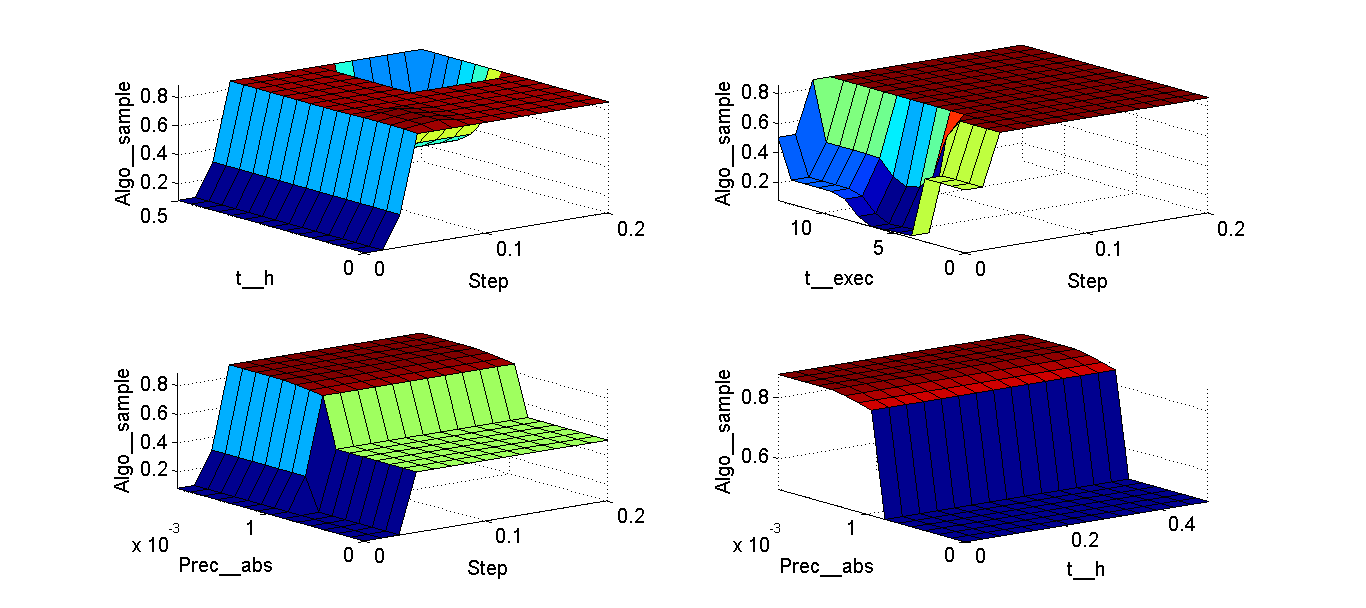}
     \caption{Response surface}
     \label{FigM}
\end{figure*}

The use of Fuzzy logic assumes an adequate knowledge of the system. It is important to analyze the output value according to the input variables. Another resource employed was that of Fig. \ref{FigM}, in which it is possible to evaluate the output according to the behavior of two of the inputs.
\section{Method Application}
\begin{figure*}[h]
     \centering
     \includegraphics[width=.9\columnwidth]{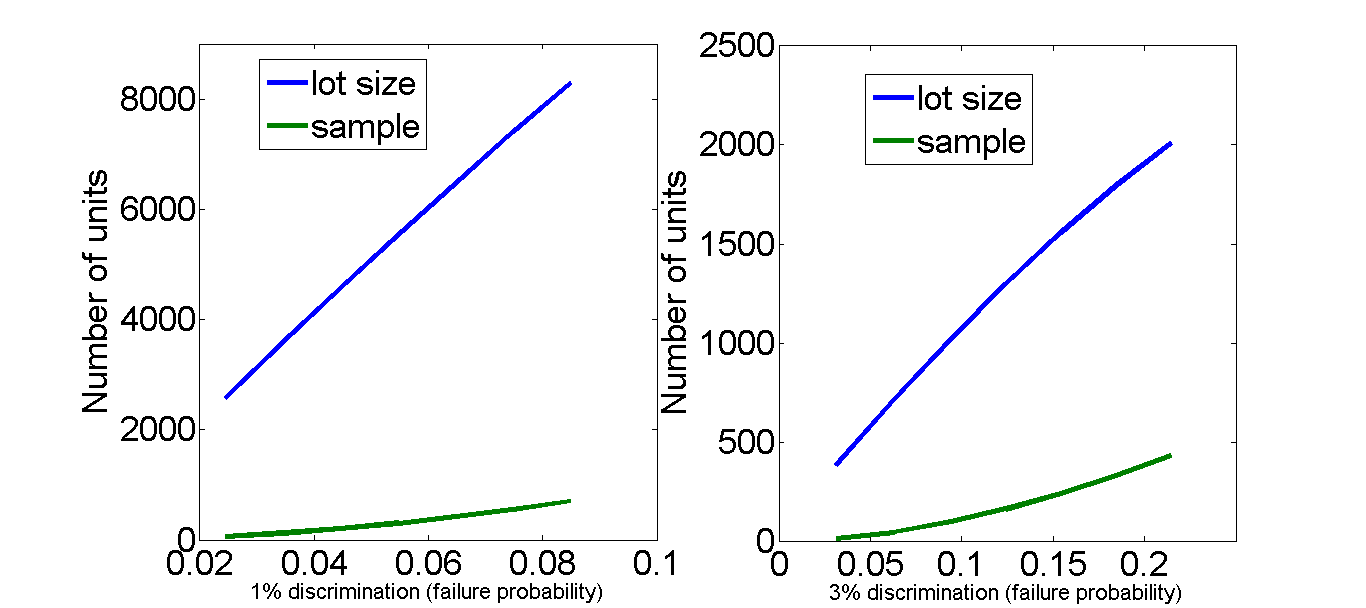}
     \caption{Lot size, number of rejections as a function of threshold}
     \label{FigN}
\end{figure*}

\begin{figure*}[htb]
     \centering
     \includegraphics[width=1.0\columnwidth]{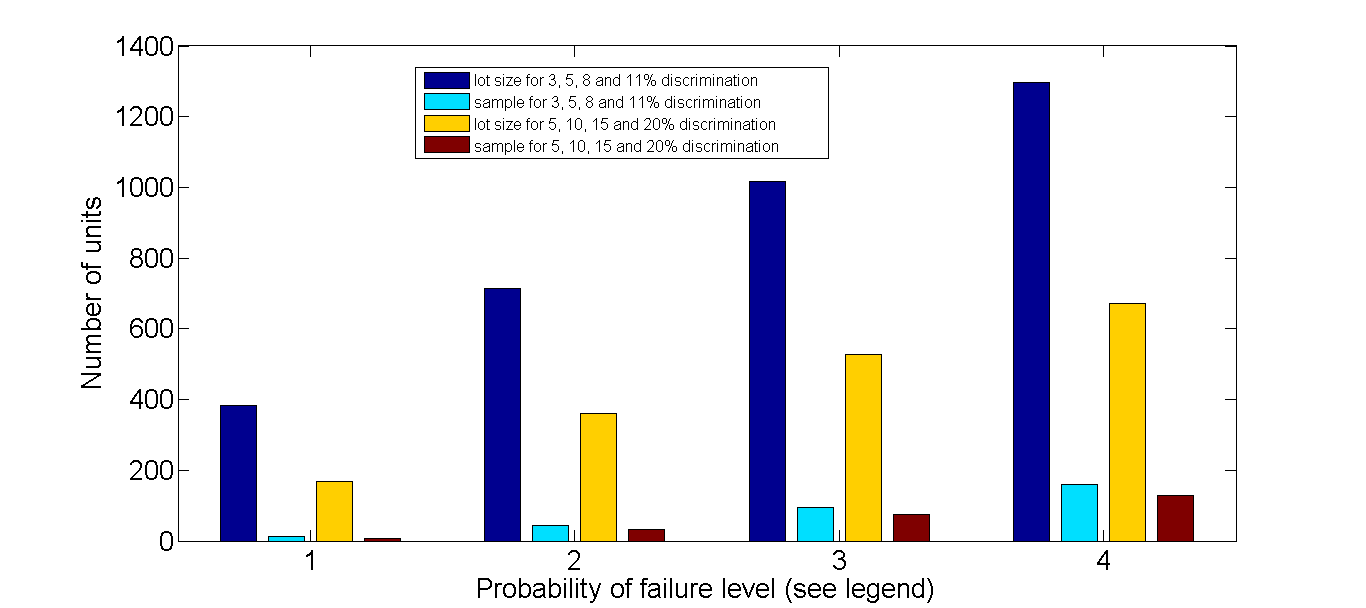}
     \caption{Lot size and number of rejections for several levels}
     \label{Fig5}
\end{figure*}

The following are steps for evaluating various Bernoulli trials sizes, and probability requirements. Using a more pragmatic approach, without taking precision into account, we use the algorithm Bin (instead of Poiss) in the first line and in the other lines Norm\_I (instead of Norm\_N). Regarding application, for example, in the first line of Table \ref{Tab2}, for step = 3 \% and with 250 Bernoulli trials: we begin with probability of failure = 0\% and we analyze the failure level between 0 and 3 \%. If there are less than 4 failures, it is assumed that the error is less than 1.44 \%. If it is larger than 4, we proceed with the test to the next level (between 3 and 6\%). Eventually we should stop if the recommended level is less than 3\%. At any tested range, if the successive failure threshold $r$ exceeds the predicted value (e.g., in the case of $p_1$ = 3\%, it cannot exceed 4 successive failures as in Fig. \ref {FigH}, you should already pass next level assessment).
\begin{center}
\begin{table}[h]
\centering
\caption{{\bf Bernoulli trials, rejection number and threshold}
 }
\label{Tab2}
\begin{tabular}{|c|c|c|c|c|c|c|c|c|c|c|c|}
\hline
\multicolumn{12}{|c|}{{\bf Values for probability steps}}\\
\hline\hline
\multicolumn{4}{|c|}{\bf step = 1\%} & \multicolumn{4}{|c|}{\bf step = 3\%} & \multicolumn{4}{|c|}{\bf step = 5\%}\\
\hline \hline
 $n$ & $c$ &  $t_h$ & $r$ &  $n$ &  $c$ &  $t_h$ & $r$ & $n$ & $c$ & $t_h$ & $r$ \\
\hline
650 &   3 & 0.0050  & 3 & 250 & 4 & 0.0143 &4 &  150 & 4 & 0.0238 &5 \\
\hline
1513    &  21  & 0.0139 &4 & 495  & 21 & 0.0424 &5 & 288 & 20 & 0.0694 &6 \\
\hline
2594    &  64  & 0.0247 &4 & 815  & 60 & 0.0736 &6 & 463 & 57 & 0.1231 &8 \\
\hline
3543     &  123  & 0.0347 & 5& 1109 & 115 & 0.1037 &7 &  615 & 107 & 0.1740 & 9\\
\hline
4517    &  202  & 0.0447 & 5& 1381 & 185 & 0.1340   &8 &   744 & 167 & 0.2245 & 10\\
\hline
5469 & 299 & 0.0547 & 5 & 1632 & 268 &0.1642 & 8& 852 &  234 & 0.2746 & 12\\
\hline 
6399   & 414  & 0.0647 & 6 & 1860  & 362  & 0.1946  & 9 & 938 & 304 & 0.3241 &13 \\
\hline
7308   & 547  & 0.0748 & 6 & 2067 & 464 & 0.2245 &  10 &   1002 & 375 & 0.3743 & 15\\
\hline
\hline
 \multicolumn{12}{|c|}{Used $\epsilon_B=0.0019$ and  $\epsilon_I=10^{-4}$}\\
\hline

\end{tabular}
\end{table}
\end{center}
If so, continuing with the example of step = 3 \%, we should proceed to the second line: testing up to 495 units, but we must have a maximum of 21 failures to ensure a percentage of failures less than or equal to 4.24 \%.
Figure \ref{FigN} shows the evolution of the lot size (Bernoulli trials) compared to the number of failures for steps of 1 \% and 3 \%. If we want more precision about the true failure rate, the lot size must undoubtedly be larger. However, when we move from one level to another, the count will continue and the inspection work already done remains valid. \\
The number of samples, besides being a function of the desired level of precision, is also a function of the sequence of steps to arrive at the batch failure probability value. However,
Sample values are cumulative and the work of the previous steps is not lost, as shown in (Fig. \ref{Fig5}) for steps of 3 \% and 5 \%.

\section{Observations and Conclusions}

There are situations which is possible to know (or roughly estimate) the probability distributions and/or the defect rates associated with the lots: components, materials, processes, or systems under inspection. In such cases it is possible to assess the occurrence of failures or the fulfillment of requirements by more appropriate procedures chosen on a case by case basis. When you have only the information that a test will be “successful” or “failure,” the approach by the DHT can have some advantages. as well as controlling and minimizing the probability of Type I and Type II errors occurring. With the two probabilities side by side, the Type I error for the lower probability distribution and the Type II error for the higher probability distribution will be estimated. The association of the DHT with the  Successive Failures Limit further improves its decision-making power. This is because if the observations were well out of range under inspection (under percentage), the SFL threshold quickly rejects the null hypothesis. Specifically, where there is no defined (nor existing) requirement for the probability of acceptance or rejection, the DHT can be applied continuously to find out the percentage of components, processes, or systems defects that are still unknown. As general recommendations and comments we can say:
\begin{itemize}
\item Continuous distributions should be used whenever $n.p>$ 5;
\item In situations where the value $p = 0$ needs to be included, the best approximation for requiring less testing, although less accurate than Poiss, is the Binomial distribution itself (Bin);
\item In general, with $n.p>5$, the Norm\_I approach, for simplicity, should be chosen;
\item In situations where accuracy is very important, or where there is not much information about the process, the Norm\_N approach should be used because it always produces a response (even when the response is invalid, e.g, $np;<5$). However, for real time applications we should look to execution time $t_{exec}$;
\item Due to the fact that this method will be applied in very general situations and with many observations, the use of SFL (Successive Failures Limit) is essential since it allows you to promptly reject a level for $p$ and start evaluating another level;
\item Unlike our proposal, sometimes the procedure assumes a fixed sample size (constant) for each batch. However, this can be easily adapted so that the value of $n$ is fixed and the value of $p$ is determined according to the value of $c$ obtained in the sample.
\end{itemize}

This work can be general enough to consider the most varied types of application. Other situations may require that the error associated with the producer be different (greater or less) from the error associated with the consumer, which may result in slightly different results than those made with $\alpha = \beta = 5 \% $. Another example, in cases where the result is dependent on $t_{exec}$ (hardware) or the number of iterations ($iter$).  Future work may include a Fuzzy control, for example, which, based on basic rules, can indicate the most appropriate method for each case or by including linguistic information about the true vale of $p$. 



\bibliographystyle{unsrtnat}
\clearpage
\bibliography{10}

\textsc{Anscombe , F. J.}  (1953). \emph{Sequential Estimation}, Journal of the Royal Statistical Society XV (1) 1209–-1215.

\textsc{Faculty Washington} (2017) \emph{Multiple Nonlinear Equations using the Newton-Raphson Method}, acessed 1n 01/04/2017.

\textsc{PennState - Eberly College of Science} (2019) \textit{Approximations for Discrete Distributions}.
acessed 1n 01/24/2019, https://online.stat.psu.edu/stat414/node/179/.

\textsc{Feller, W.} (1968)
\textit{An Application to Probability Theory and Its Applications}.
John Wiley \& Sons, third edition, Chapter XIII, Recurrent Events, Renewal Theory, pg. 324.

\textsc{flestad, B.} (2006)\textit{Newton’s method for systems of non--linear equations}, acessed in 01/04/2017.

\textsc{Hashemi, F. S., Pasupathy, R.}(2012) \textit{Averaging and Derivative Estimation Within Stochastic Estimation Algorithms}, Proceedings of the  Winter Simulation Conference, WSC12.

\textsc{Hashemi, F. S., Ghosh, S., Pasupathy,  R.} (2014) \textit{On Adaptive Sampling Rules for Stochastic Recursions}, Proceedings of the  Winter Simulation Conference, WSC14.

\textsc{Istratescu. V. I.} (1981) \textit{Fixed Point Theory, An Introduction}, D. Reidel Publishing Company, The Netherlands, chapter 7.

\textsc{Louren{\c c}o Filho, R. de C. B.} (1978), \textit{Controle Estatistico de Qualidade}, Livros Tecnicos e Cientificos Editora S. A.

\textsc{Ng, S. W., Lee, Y. S.}(2000) \textit{Variable Dimension, Newton-Raphson Method}, IEEE Transactions on Circuits and Systems –- I: Fundamental Theory and Applications Vol. 47, Nº 6.

\textsc{Jamkhaneh, E, B., Sadeghpour-Gildeh, G., Yari,} .(2009) \textit{Acceptance Single Sampling Plan with fuzzy parameter with The Using of Poisson Distribution}, World Academy of Science, Engineering and Technology 25, pg. 1017--1021.

\end{document}